\documentclass[12pt]{article}
\usepackage{amsmath}
\usepackage{graphicx}
\usepackage{float}
\begin{document}

\title{{\bf Time Dependence\\ of\\ Hawking Radiation Entropy}}

\author{
Don N. Page
\thanks{Internet address:
profdonpage@gmail.com}
\\
Department of Physics\\
4-183 CCIS\\
University of Alberta\\
Edmonton, Alberta T6G 2E1\\
Canada
}

\date{2013 August 9}

\maketitle
\large
\begin{abstract}
\baselineskip 15 pt

If a black hole starts in a pure quantum state and evaporates completely by a unitary process, the von Neumann entropy of the Hawking radiation initially increases and then decreases back to zero when the black hole has disappeared.  Here numerical results are given for an approximation to the time dependence of the radiation entropy under an assumption of fast scrambling, for large nonrotating black holes that emit essentially only photons and gravitons.  The maximum of the von Neumann entropy then occurs after about 53.81\% of the evaporation time, when the black hole has lost about 40.25\% of its original Bekenstein-Hawking (BH) entropy (an upper bound for its von Neumann entropy) and then has a BH entropy that equals the entropy in the radiation, which is about 59.75\% of the original BH entropy $4\pi M_0^2$, or about $7.509 M_0^2 \approx 6.268\times 10^{76}(M_0/M_\odot)^2$, using my 1976 calculations that the photon and graviton emission process into empty space gives about 1.4847 times the BH entropy loss of the black hole. Results are also given for black holes in initially impure states.  If the black hole starts in a maximally mixed state, the von Neumann entropy of the Hawking radiation increases from zero up to a maximum of about 119.51\% of the original BH entropy, or about $15.018 M_0^2 \approx 1.254\times 10^{77}(M_0/M_\odot)^2$, and then decreases back down to $4\pi M_0^2 = 1.049\times 10^{77}(M_0/M_\odot)^2$.

\end{abstract}


\baselineskip 22 pt

\newpage

Interest in black hole information \cite{Hawking:1976ra,Page:1979tc,Wald:1980nm,Page:1982fk,Hawking:1982dj, Page:1982ue,AlvarezGaume:1982fy,Gross:1983mq,Strominger:1983ns,'tHooft:1984re, Bowick:1986km,Carlitz:1986ng,Hawking:1987mz,Hawking:1988wm,'tHooft:1990fr, Schwarz:1991mv,Harvey:1992xk,Preskill:1992tc,Banks:1992is,Bekenstein:1993bg, Wilczek:1993jn,Danielsson:1993um,Page:1993up,Susskind:1993if,Page:1993wv, Susskind:1993ki,Susskind:1993mu,Stephens:1993an,Moffat:1993ep,Fiola:1994ir, Polchinski:1994zs,Strominger:1994ey,Page:1994dx,Frolov:1994ek,Banks:1994ph, Danielsson:1994qt,Strominger:1994tn,Lowe:1995pu,Polchinski:1995ta, Itzhaki:1995tc,Nastase:1996ue,Callan:1996dv,'tHooft:1996tq,Horowitz:1996tx, Englert:1996az,Kay:1998vv,Lowe:1999pk,Hajicek:1999dg,Bigatti:1999dp, 'tHooft:1999bw,Tipler:2000zy,Peet:2000hn,Das:2000su,Maldacena:2001kr, Susskind:2002ri,Srednicki:2002fe,ShalytMargolin:2002wa,Qi:2002uy, Horowitz:2003he,Bekenstein:2003dt,Gottesman:2003up,Giddings:2004ud,Das:2004cs, Thorlacius:2004nn,Parikh:2004ih,Solodukhin:2004rv,Valentini:2004ep, Flambaum:2004ru,Rabinowitz:2004mv,Solodukhin:2005qy,Ge:2005av,Russo:2005aw, Barbon:2005jr,Medved:2005vw,Zeh:2005ka,Farley:2005px,Hawking:2005kf, Pullin:2005xr,Einhorn:2005bi,Lowe:2006xm,Braunstein:2006sj,Giddings:2006vu, Ahn:2006uc,Giddings:2006sj,Ahn:2006zk,Giddings:2006be,Ahn:2006wi,Hsu:2006pe, Brustein:2006wp,Bena:2007kg,Giddings:2007ie,Srikanth:2007vz,Giddings:2007pj, Nikolic:2007mu,Hayden:2007cs,Herrera:2007ye,Lin:2007mu,Chen:2007zzu, Barcelo:2007yk,Ahn:2008zf,Mathur:2008wi,Bambi:2008tr,Hsu:2008yi, Skenderis:2008qn,Nikolic:2008jc,Mathur:2008kg,Pankovic:2008yh,Iizuka:2008eb, Hossenfelder:2009xq,Zhang:2009jn,Hsu:2009ve,Liberati:2009ak,Nikolic:2009ju, Chen:2009ut,Mathur:2009zs,Strominger:2009aj,Zhang:2009td,Yeom:2009zp, Dong:2009fn,Mathur:2009hf,Shiokawa:2009ck,Giddings:2009ae,Yeom:2009mn, Nikolic:2009hr,Sakalli:2010yy,Dragan:2010kz,Hsu:2010sb,Mathur:2010dg, Giddings:2010pp,Ahn:2010xg,Ahn:2010yp,Pappas:2010fq,Englert:2010cg, Zhang:2010aq,Mathur:2010kx,Liu:2010ai,Nikolic:2011rf,Mathur:2011wg, Balasubramanian:2011dm,Giddings:2011xs,Czech:2011wy,Mathur:2011uj, Smolin:2011ns,Giddings:2011ks,Cantcheff:2011jk,Pappas:2012sg,Giddings:2012bm, Mathur:2012np,Mathur:2012zp,Mathur:2012vb,Giddings:2012dh,Mueller:2012ss, Mathur:2012dxa,Brustein:2012jn,Kay:2012vw,Cai:2012um,Nomura:2012cx,Dvali:2012wq, Veneziano:2012yj,Faizal:2013ji}
has surged recently with the publication by Almheiri, Marolf, Polchinski, and Sully \cite{AMPS} of a provocative argument that suggests that an ``infalling observer burns up at the horizon'' of a sufficiently old black hole, so that the horizon becomes what they called a ``firewall.''  This paper has elicited a large number of responses, some of which support the firewall idea \cite{Bousso:2012as, Susskind:2012rm, Susskind:2012uw, Giveon:2012kp, Saravani:2012is}, others seem rather agnostic \cite{Bena:2012zi, Hwang:2012nn, Culetu:2012fh}, and yet others of which raise skepticism about it \cite{Nomura:2012sw, Mathur:2012jk, Chowdhury:2012vd, Banks:2012nn, Ori:2012jx, Brustein:2012jn, Hossenfelder:2012mr, Nomura:2012cx, Avery:2012tf, Larjo:2012jt, Rama:2012fm, Page:2012zc, Papadodimas:2012aq, Nomura:2012ex, Giddings:2012gc, Jacobson:2012gh}.

A central question concerns the entanglement of a black hole with the Hawking radiation \cite{Hawking:1974rv,Hawking:1974sw} that has been emitted.  In fact, it is the argument that at late times the Hawking radiation is maximally entangled with what is inside the black hole that suggests that what is just inside cannot be significantly entangled with what is just outside, and without this latter entanglement, an observer falling into the black hole should encounter high-energy radiation at the horizon that would burn up the observer \cite{AMPS}.

It would be impossible for an observer to remain outside the black hole horizon and yet to measure directly the entanglement or lack thereof across it.  However, if the black hole emission is a unitary process, the entanglement between the black hole and the earlier Hawking radiation should eventually be transferred to an entanglement between the earlier Hawking radiation and later Hawking radiation.  This should have an effect on the time dependence of the von Neumann entropy of the radiation up to some point in retarded time during the emission.  That would be a function of that time that in principle should be calculable (though not yet in practice) from a theory of quantum gravity that considers only the black hole exterior and indeed only what gets radiated to future null infinity.

The importance of this question of the time dependence of the Hawking radiation entropy has recently been emphasized by Strominger \cite{Strominger:2009aj}.  He gives a brief outline of five candidate answers to the question (bad question, information destruction, long-lived remnant, non-local remnant, and maximal information return).  Here I shall assume without proof that the fifth answer is correct and make an additional assumption of nearly maximal entanglement between the black hole and the Hawking radiation \cite{Page:1993df,Page:1993wv} (with support from the later conjecture of fast scrambling \cite{Sekino:2008he, Susskind:2011ap,Lashkari:2011yi,Edalati:2012jj,Barbon:2012zv}).  In particular, I shall assume that the von Neumann entropy of the radiation is very nearly that given by the semiclassical approximation up to the point where that equals the Bekenstein-Hawking entropy of the remaining black hole, and thereafter is very nearly the same as the Bekenstein-Hawking entropy of the shrinking hole.  Then I shall give, from my previous numerical calculations \cite{Page:1976df,Page:1976ki,Page:1977um}, an approximation to the time dependence at future null infinity of the von Neumann entropy of the Hawking radiation up to that point from a large nonrotating black hole. 

I shall consider only nonrotating uncharged (Schwarzschild) black holes that are sufficiently large (and hence at low enough Hawking temperature) that they emit essentially only massless particles, which I shall assume are only photons and gravitons.  I shall assume that all other particles (e.g., neutrinos) are massive, that the lightest massive particle (perhaps the electron neutrino) has mass $m$, and that one can neglect the radiation from a black hole into particles of mass $m$ if (using Planck units) $8\pi < m/T = 8\pi mM$, where $M$ is the mass of the black hole.  My numerical calculations \cite{Page:1977um} showed that for a neutral particle species of spin-half, the power emission for $m = M^{-1}$ is only $1.505\times 10^{-8}$ times that for spin-half particles with $m=0$, so indeed it is a good approximation to neglect the Hawking emission of particles with $Mm > 1$.

Therefore, I shall restrict attention to large black holes, with $M > m^{-1} \approx 1.336\,337\,707\times 10^{-10} M_\odot\mathrm{eV}/(mc^2)$ \cite{Beringer:1900zz}.  Any black hole of a solar mass $M_\odot$ or greater would qualify so long as the lightest massive particle has a rest mass energy greater than about 0.134 nano-electron volts, which is over ten orders of magnitude smaller than the current upper limit of 2 eV on the anti-electron-neutrino mass \cite{Beringer:1900zz} and nearly eight orders of magnitude smaller than 0.0087 eV, the square root of the smallest known difference [$(7.50\pm 0.20)\times 10^{-5}\ \mathrm{eV}^2$] of the squares of two of the three neutrino mass-energies \cite{Beringer:1900zz}.  None of these data exclude the possibility of one massless neutrino species (or of other massless particles besides photons and gravitons, provided their coupling to known particles is sufficiently weak), but without any theoretical reason to expect such massless species, it seems most likely that there are no massless particles other than photons and gravitons (for which there are strong theoretical reasons from the assumption of gauge invariance for suspecting that they are precisely massless), and no massive particles of smaller rest-mass energy than $1.34\times 10^{-10}$ eV.  If this is indeed true, as I shall assume here, black holes of solar masses and above would radiate almost entirely photons and gravitons.

My numerical calculations (with an estimated accuracy of about four or five digits; not all of the digits given below may be significant, but I give them so that the numerical error of my calculations can be found if anyone repeats the calculations to higher precision) of the Hawking particle emission rates from both nonrotating and rotating black holes \cite{Page:1976df,Page:1976ki,Page:1977um} (see also \cite{Zurek:1982zz,Page:1983ug,Page:2006my}) in Planck units showed (e.g., in Table II of \cite{Page:1976ki}, summarized in \cite{Page:1983ug} and in Eq.\ (19) of \cite{Page:2006my}) that the emission power in photons is approximately $0.000\,033\,638/M^2$ and in gravitons is close to $0.000\,003\,836/M^2$, for a total emission rate of massless radiation energy of $\alpha/M^2 \approx 0.000\,037\,474/M^2$ with
\begin{equation}
\alpha \equiv -M^2\frac{dM}{dt} \approx 0.000\,037\,474.
\label{power-coefficient}
\end{equation}

My numerical calculations also showed (e.g., summarized in \cite{Page:1983ug} and in Eq.\ (20) of \cite{Page:2006my}) that the rate of entropy emission in photons is approximately $0.001\,268\,4/M$ and in gravitons is about $0.000\,130\,0/M$, for a total emission rate of massless radiation entropy of $dS_\mathrm{rad}/dt \approx 0.001\,398\,4/M$.  These latter numbers were calculated from the former by using the results I reported in the last paragraph of the first column of the tenth page (page 3269) of \cite{Page:1976ki}:

``There is still some entropy produced by the partial scattering off the gravitational potential barrier surrounding the hole, but outside the supperradiant regime this can only partially cancel the entropy flow out of the hole and serves in effect to increase the entropy emitted to the surrounding region for a given entropy loss by the hole.  For example, numerical calculations for a nonrotating hole show that the emission of $s = \frac{1}{2}$ particles into empty space increases the external entropy by 1.6391 times the entropy drawn out of the hole, $s = 1$ particles increase it by a factor of 1.5003, $s = 2$ particles by 1.3481, and the canonical combination of species [the set of species known in 1976 with masses less than 20 MeV: gravitons, photons, electrons and muon neutrinos with one helicity each, electrons, and the corresponding antileptons, all taken to have $Mm \ll 1$ for this numerical result] gives 1.6233 times as much entropy in radiation as the entropy decrease of the hole.''

When one uses the fact that the Bekenstein-Hawking semiclassical entropy \cite{Bekenstein:1972tm,Bekenstein:1973ur,Hawking:1974sw} of a nonrotating uncharged black hole is, in Planck units, $S_\mathrm{BH} = 4\pi M^2$ \cite{Hawking:1974sw}, the rate of entropy decrease of the hole is $8\pi\alpha/M$.  From just photon emission, one would get (cf.\ \cite{Page:1983ug} and Eq.\ (20) of \cite{Page:2006my}) $0.000\,845\,41/M$, and from just graviton emission one would get $0.000\,096\,41/M$, for a total rate of decrease of the Bekenstein-Hawking black hole entropy of $-dS_\mathrm{BH}/dt = 8\pi\alpha/M \approx 0.000\,941\,82/M$.  This then gives the ratio by which the increase in the coarse-grained entropy of the Hawking radiation (e.g., ignoring its entanglement with the black hole) is greater than the decrease in the coarse-grained Bekenstein-Hawking entropy of the black hole as
\begin{equation}
\beta \equiv \frac{dS_\mathrm{rad}/dt}{-dS_\mathrm{BH}/dt} \approx 1.484\,72;
\label{entropy-coefficient}
\end{equation}
the last one or two digits probably are not significant but are given as a challenge for someone to do a higher-precision calculation to find the error. (This constant $\beta$ should not be confused with the $\beta$ given in Table II of \cite{Page:1976ki}, which gave a measure of the angular momentum emission of very slowly rotating black holes.)

From the fact that the mass of a large nonrotating black hole that emits almost entirely just massless photons and gravitons decreases according to $dM/dt = -\alpha/M^2$, one can get that the time evolution of the black hole mass is
\begin{equation}
M(t) = (M_0^3 - 3\alpha t)^{1/3} = [3\alpha(t_\mathrm{decay} - t)]^{1/3}
     = M_0(1 - t/t_\mathrm{decay})^{1/3},
\label{mass-evolution}
\end{equation}
where $M_0$ is the initial mass of the black hole at time $t=0$ and where
\begin{equation}
t_\mathrm{decay} = \gamma M_0^3 \equiv \frac{1}{3\alpha}M_0^3 \approx 8895 M_0^3 \approx 1.159\times 10^{67}\left(\frac{M_0}{M_\odot}\right)^3\mathrm{yr}
\label{decay-time}
\end{equation}
is what the decay time would be of a nonrotating uncharged black hole of initial mass $M_0$ if only massless photons and gravitons were emitted, with
\begin{equation}
\gamma \equiv \frac{1}{3\alpha} \equiv \frac{1}{-3M^2 dM/dt} \approx 8895.
\label{decay-coefficient}
\end{equation}

Assuming that the lightest massive particle has mass $m$, the emission of massive particles would reduce the actual decay time by an amount of the order of $1/m^3$ in Planck units, e.g., by a time of the order of $10^{33}(\mathrm{eV}/m)^3$ years, which for $m \stackrel{<}{\sim} 0.1$ GeV is greater than the age of the universe but which for $m \gg 10^{-10}$ eV (which I am assuming) is much less than the decay time $t_\mathrm{decay} \sim 10^{67}$ years for a solar-mass black hole, so in this paper I shall ignore any such correction to the black hole decay time.  Another correction that I shall mostly ignore but shall return to later is the uncertainty of the decay time from the stochastic nature of the Hawking emission of very roughly $N \sim M_0^2$ particles, which would lead to fluctuations in the total decay time of an ensemble of black holes of initial mass $M_0$ that would be expected to be of the order of $\delta t \sim t_\mathrm{decay}/\sqrt{N} \sim t_\mathrm{decay}/M_0 \sim M_0^2 \sim 10^{25} (M_0/M_\odot)^2$ years.  These fluctuations in the total decay time have magnitudes that are comparable (with $c=1$) to the fluctuations in the position of the final decay of the black hole in the center-of-mass frame of the initial black hole, $\delta x \sim M_0^2 \sim 10^{25} (M_0/M_\odot)^2$ light-years \cite{Page:1979tc,Nomura:2012cx,Nomura:2012ex}.

The semiclassical approximation for the black hole emission \cite{Hawking:1974rv,Hawking:1974sw,Hawking:1976ra} would then give the time-dependence of the coarse-grained entropy of a decaying large nonrotating uncharged black hole and of the emitted radiation as
\begin{equation}
\tilde{S}_\mathrm{BH}(t) 
= 4\pi M_0^2\left(1-\frac{t}{\gamma M_0^3}\right)^{2/3}
\approx 4\pi M_0^2\left(1-\frac{t}{8895 M_0^3}\right)^{2/3},
\label{bh-macro-entropy-evolution}
\end{equation}
\begin{equation}
\begin{split}
\tilde{S}_\mathrm{rad}(t) 
&= 4\pi\beta M_0^2\left[1-\left(1-\frac{t}{\gamma M_0^3}\right)^{2/3}\right]
\\
&\approx 4\pi(1.4847) M_0^2\left[1-\left(1-\frac{t}{8895 M_0^3}\right)^{2/3} \right].
\label{rad-macro-entropy-evolution}
\end{split}
\end{equation}

Here I have put tildes on the expressions for these entropies to emphasize that they are the entropies obtained from semiclassical approximations.  Ultimately we would like instead to obtain von Neumann entropies of black hole and radiation subsystems ($S_\mathrm{vN}(t)$) of the entire universe, which for simplicity I shall assume is in a pure quantum state with zero von Neumann entropy.  Also for simplicity in the first part of this paper I shall assume that one is considering a black hole that has formed in a pure quantum state, which implies that it has no quantum entanglement with the rest of the universe.  Of course, these are unrealistic assumptions, but they might give a reasonably good approximation if one focuses on one pure component of the quantum state of the universe in which the black hole forms from the gravitational collapse of a star with a von Neumann entropy of the order of the number of particles in the star, say $S \sim 10^{57}$, which is much smaller than the Bekenstein-Hawking entropy of a stellar mass black hole that is at least $S_\mathrm{BH} = 4\pi M_\odot^2 \sim 10^{77}$.  In other words, in comparison with the von Neumann entropies to be discussed that are at least of the order of $10^{77}$, I shall neglect the much smaller ordinary stellar entropies that are of the order of $10^{57}$.  In the latter part of this paper I shall also consider the more hypothetical cases in which the initial von Neumann entropy of the black hole is a significant fraction $f$ of its initial semiclassical entropy.

I am also not making a sharp boundary between the black hole and the rest of the universe, such as the horizon with a Planck-scale cutoff, which would lead to an entanglement entropy between the black hole inside the boundary and the quantum fields just outside the boundary that would be of the order of the area of the horizon in Planck units \cite{Sorkin:1985bu,Bombelli:1986rw,Srednicki:1993im}.  Instead, I am envisaging some sort of fuzzy boundary, say between roughly $r \sim 3M$ and $r \sim 6M$, so that the entanglement entropy between the black hole region inside the black hole and nearby region (say for $r < 3M$) and any region outside the fuzzy boundary (say for $6M < r < 12M$) that does not extend out to where the bulk of the Hawking radiation might be expected to be is only a few units (e.g., roughly the number of Hawking quanta that might be expected to be at one time propagating between $r = 6M$ and $r = 12M$).  I do not know how to specify such a fuzzy bondary in a precise way, but I do assume some prescription in which the von Neumann entropy of the black hole region inside this fuzzy boundary can be much less than the coarse-grained Bekenstein-Hawking entropy $A/4$, such as one might expect if, say, the black hole forms from a star with coarse-grained entropy much less than $A/4$.

I shall take the semiclassical entropies to be upper limits on the von Neumann entropies of the corresponding subsystems with the same macroscopic parameters.  That is, I shall first assume that the semiclassical Bekenstein-Hawking entropy of a nonrotating uncharged black hole is a good approximation for the maximum von Neumann entropy of a black hole of the same energy, at least if one neglects entropy associated with the location and/or motion of the black hole in a space sufficiently large that this could in principle rival the Bekenstein-Hawking entropy.  Secondly, I shall assume that the semiclassical entropy calculated for the Hawking radiation is the maximum von Neumann entropy for radiation with the same expectation values of the numbers of particles in each of the modes.  This second assumption is supported by the fact that indeed the semiclassical approximation does give a thermal density matrix for each radiation mode, and no entanglement between different modes, though because of the black-hole greybody factors the temperature varies from mode to mode.

Now under the assumptions that the black hole starts in essentially a pure state (von Neumann entropy much less than the Bekenstein-Hawking entropy), that the Hawking evaporation is a unitary process, and that we can neglect the interaction with other systems (e.g., no large external source of entropy impinging upon the black hole or interacting with the outgoing radiation), the von Neumann entropy of the evaporating black hole equals that of the Hawking radiation that has been emitted.  When combined with the assumptions of the previous paragraph, we reach the conclusion that at retarded time $t$, the von Neumann entropy of the black hole or of the Hawking radiation propagating outward before that retarded time cannot exceed the minimum of the two semiclassical entropies given above, $\tilde{S}_\mathrm{BH}(t)$ and $\tilde{S}_\mathrm{rad}(t)$.  (We also assume a fuzzy boundary between what is before and what is after the retarded time $t$ to avoid large entanglement entropies between the parts of the Hawking radiation on opposite sides to, but very near, that boundary.)

\newpage

Next, based upon what is suggested by the results of \cite{Page:1993df,Page:1993wv,Sekino:2008he,Susskind:2011ap,Lashkari:2011yi, Edalati:2012jj,Barbon:2012zv}), as discussed above, I shall assume that the actual von Neumann entropy of the Hawking radiation up to retarded time $t$ (which is also that of the remaining black hole, though from now on I shall focus on the radiation entropy as potentially being more easily measured) is very near to its maximal allowed value that is the minimum of $\tilde{S}_\mathrm{BH}(t)$ and $\tilde{S}_\mathrm{rad}(t)$.  One might summarize this assumption as a special case of what I might call the {\it {Conjectured Anorexic Triangle Hypothesis (CATH)}} \cite{Cathy}:  

{\bf An entropy triangular inequality tends to be saturated.}

$\tilde{S}_\mathrm{BH}(t)$ decreases monotonically with time for $0 < t < t_\mathrm{decay} = \gamma M_0^3 \approx 8895 M_0^3$, and $\tilde{S}_\mathrm{rad}(t)$ increases monotonically with time.  The two values cross at the time
\begin{equation}
\begin{split}
t_\ast & = \left[1-\left(\frac{\beta}{\beta+1}\right)^{3/2}\right]t_\mathrm{decay} \equiv \epsilon t_\mathrm{decay} \\
& \approx 0.53810\, t_\mathrm{decay} \approx 4786 M_0^3 
\approx 6.236\!\times\!10^{66}\!\left(\frac{M_0}{M_\odot}\right)^3\!\mathrm{yr},
\label{Page-time}
\end{split}
\end{equation}
with 
\begin{equation}
\epsilon \equiv
 \left[1-\left(\frac{\beta}{\beta+1}\right)^{3/2}\right] \approx 0.53810,
\label{Page-time-fraction}
\end{equation}
at which time the mass of the black hole is
\begin{equation}
M_\ast = \left(\frac{\beta}{\beta+1}\right)^{1/2}\approx 0.77301 M_0, 
\label{Page-mass}
\end{equation}
and the entropy of the radiation and of the black hole is
\begin{equation}
\begin{split}
S_\ast \! &\equiv \! S_\mathrm{vN}(t_\ast) \! = \! 
\tilde{S}_\mathrm{BH}(t_\ast) \! = \! \tilde{S}_\mathrm{rad}(t_\ast)
\! = \! \left(\frac{\beta}{\beta+1}\right)4\pi M_0^2
\! \approx \! 0.59754 \,\tilde{S}_\mathrm{BH}(0)
\\
&= 0.59754(4\pi M_0^2) \approx 7.5089 M_0^2
 \approx 6.268\times 10^{76} (M_0/M_\odot)^2. 
\label{Page-entropy}
\end{split}
\end{equation}

\newpage

This time, $t_\ast \approx 0.5381\, t_\mathrm{decay}$, under the assumptions being made in this paper, would be the retarded time at which the von Neumann entropy of the Hawking radiation from a large nonrotating uncharged black hole, which was initially in essentially a pure quantum state, reaches its peak value and thereafter decreases.  It is what I think is intended to be meant by what is called the `Page time' in the literature \cite{Czech:2011wy,AMPS,Bousso:2012as,Nomura:2012sw, Chowdhury:2012vd, Susskind:2012rm,Giveon:2012kp,Banks:2012nn,Susskind:2012uw,Nomura:2012cx, Avery:2012tf,Larjo:2012jt,Nomura:2012ex,Giddings:2012gc,Dvali:2012wq, Veneziano:2012yj}, often citing \cite{Page:1993df,Page:1993wv},
though sometimes \cite{AMPS,Bousso:2012as,Nomura:2012sw,Susskind:2012rm, Susskind:2012uw,Nomura:2012cx, Nomura:2012ex} it is somewhat inaccurately described as something equivalent to the time ``when the black hole has emitted half of its initial Bekenstein-Hawking entropy'' \cite{AMPS}, which would be correct if $\beta$, the ratio of the semiclassical entropy emitted in Hawking radiation to the decrease in the Bekenstein-Hawking semiclassical entropy of the black hole, were unity, but it is somewhat different when $\beta \neq 1$, as for the emission of photons and gravitons from a nonrotating uncharged black hole that gives $\beta \approx 1.4847$.  In particular, the time at which the black hole area has decreased to half its original value is $t_{1/2} = (1-2^{-3/2})t_\mathrm{decay} \approx 0.6464\, t_\mathrm{decay} \approx 1.201\, t_\ast$.

As we can see above, under the CATH assumptions that lead to the von Neumann entropy of the radiation being (very nearly) the minimum of the semiclassical entropies of the black hole and of the Hawking radiation, the time $t_\ast$ is actually about 53.81\% of the black hole lifetime (or about 83.24\% of the time until half the area and half the Bekenstein-Hawking entropy are lost), but on the other hand at this time the black hole has lost only about 40.25\% of its original Bekenstein-Hawking entropy.  The excess entropy in the radiation over what is lost from the black hole comes from the semiclassical entropy generated in the emission process, which is a nonequilibrium transfer of energy from a black hole of positive temperature to originally empty space of zero temperature.  That is, the loss of about 40.25\% of the original black hole entropy generates radiation of $\beta \approx 1.4847$ times as much, or about 59.75\% of the original black hole entropy, equaling the remaining black hole Bekenstein-Hawking entropy.

Now under the assumptions above, we can say that the von Neumann entropy of the Hawking radiation, $S_\mathrm{vN}(t)$, as a function of the retarded time $t$, is very nearly the semiclassical radiation entropy $\tilde{S}_\mathrm{rad}(t)$   for $t < t_\ast$ and is very nearly the Bekenstein-Hawking semiclassical black hole entropy $\tilde{S}_\mathrm{BH}(t)$ for $t > t_\ast$.  Using the Heaviside step function $\theta(x)$, the von Neumann entropy of the Hawking radiation from a large nonrotating uncharged black hole may be written as
\begin{equation}
\begin{split}
S_\mathrm{vN}(t) \approx {} &
4\pi\beta M_0^2\left[1-\left(1-\frac{t}{t_\mathrm{decay}}\right)^{2/3}\right]
\theta(t_\ast - t) \\ 
& +4\pi M_0^2\left(1-\frac{t}{t_\mathrm{decay}}\right)^{2/3}\theta(t - t_\ast)
\\
\approx {} &
4\pi(1.4847)M_0^2\left[1-\left(1-\frac{t}{8895 M_0^3}\right)^{2/3}\right]
\theta(4786M_0^3 - t)
\\
& +4\pi M_0^2\left(1-\frac{t}{8895 M_0^3}\right)^{2/3}
\theta(t - 4786 M_0^3). 
\label{rad-von-Neumann-entropy-evolution}
\end{split}
\end{equation}
See Figure 1 for a plot of this entropy versus retarded time.

\begin{figure}[H]
\centering
\includegraphics[scale=.75]
{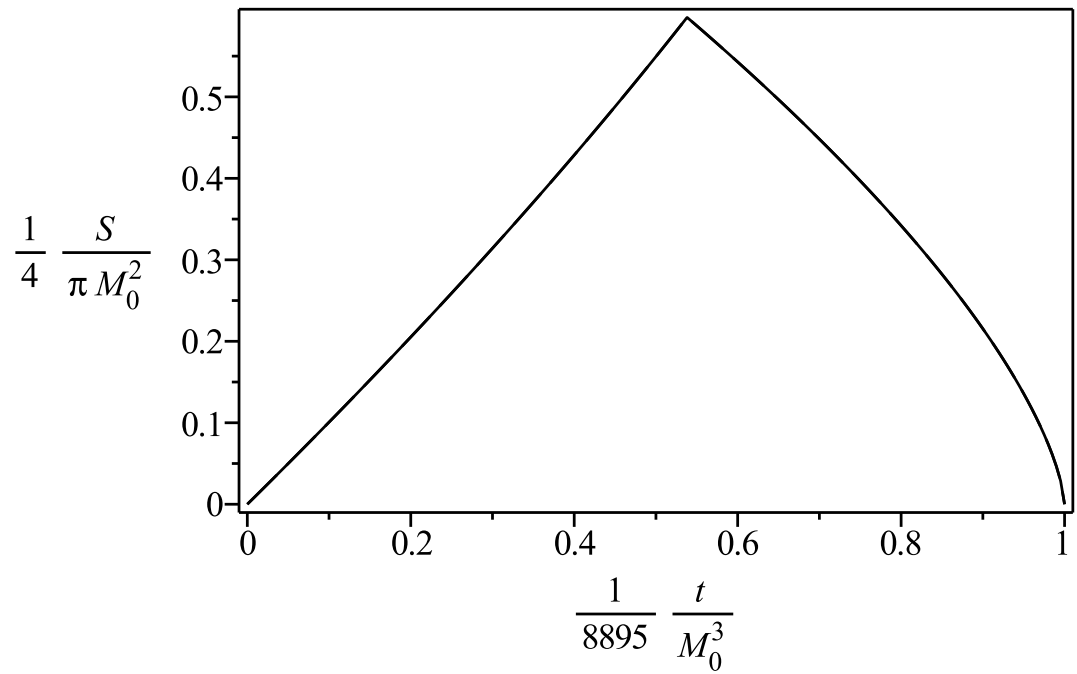}
\caption{Hawking Radiation Entropy vs.\ Time for an Initially Pure Black Hole}
\end{figure}

\newpage
\baselineskip 21 pt

Because of the fluctuations in the black hole emission rate that make the time for a black hole to evaporate down from an initial mass of $M_0$ to $M_\ast \approx 0.7730 M_0$ uncertain by an amount of the order of $M_0^2$ in Planck units, one would expect the peak in the von Neumann entropy function $S_\mathrm{vN}(t)$ at $t = t_\ast$ to be rounded off over a time scale of the order of $M_0^2$, which of course for large black holes is very much smaller than $t_\ast \approx 4786 M_0^3$, smaller by a factor of the order of $1/M_0$ that is $\sim 10^{-38}$ for a solar mass black hole.  Furthermore, if the von Neumann entropy of the radiation up to retarded time $t$ is evaluated at future null infinity up to its intersection with a null cone moving outward from the center of mass position of the total system of the black hole plus radiation in the center-of-momentum frame of that entire system, the fluctuations of the black hole position from that center of mass that are of the order of $M_0^2$ \cite{Page:1979tc,Nomura:2012cx,Nomura:2012ex} would also be expected to contribute to a rounding off of the entropy function $S_\mathrm{vN}(t)$ by a comparable amount.  Since $|dS_\mathrm{vN}(t)/dt| \sim 1/M_0$ on either side of the idealized sharp peak (ignoring this rounding off), the rounding off over a time of the order of $M_0^2$ would be expected to reduce the maximum value of the peak by an amount $\sim M_0$.  Therefore, I would expect the peak of the von Neumann entropy of the Hawking radiation actually to be $S_\ast \approx [4\pi\beta/(\beta+1)]M_0^2 - O(M_0) \approx 0.59754 \,\tilde{S}_\mathrm{BH}(0) = 0.59754(4\pi M_0^2) \approx 7.5089 M_0^2 \approx 6.268\times 10^{76} (M_0/M_\odot)^2$, with a correction (beyond the numerical errors of the coefficients) only of the order of $M_0 \sim 10^{38}(M_0/M_\odot)$.

One might think that the fluctuations of the black hole position would greatly increase the von Neumann entropy of the Hawking radiation, since if one uses radiation wavepacket modes of definite angular momenta relative to the center of mass of the entire system, the deviation of the black hole from this center of mass when it emits a mode will cause a particle in the mode to have a typical angular momentum of the order of the frequency of the mode, $\sim 1/M \sim1/M_0$, multiplied by the deviation of the black hole position from the center of mass, $\sim M_0^2$, for an angular momentum $j\sim M_0$ about the center of mass.  For each time period $\sim M \sim M_0$ over which roughly one particle is emitted by the black hole, the number of angular momentum modes into which the particle could be emitted is $\sim j^2 \sim M_0^2$.  Therefore, the total emission of $\sim M_0^2$ Hawking particles would be spread over $\sim M_0^4$ modes quasi-localized in both time and angular momentum, with each mode having a probability of being occupied that is $p \sim M_0^{-2}$ and hence a von Neumann entropy per mode $S_\mathrm{mode} \approx -p\ln{p} \sim M_0^{-2} \ln{M_0}$.  If one then sums this von Neumann entropy per mode over the $\sim M_0^4$ modes that each have roughly this entropy, one would get a total entropy in the radiation of $S_\mathrm{rad\ with\ flucts} \sim j^2 S_\mathrm{mode} \sim M_0^2 \ln{M_0}$, roughly a factor of the order of $\ln{M_0} \sim 87 + \ln{(M_0/M_\odot)}$ times the Bekenstein-Hawking semiclassical entropy of the original black hole of mass $M_0$.

However, this entropy calculated for the Hawking radiation coming from a black hole, with fluctuations in its position $\sim M_0^2$ from fluctuations in the momenta of the previously emitted quanta, is a highly coarse-grained entropy, ignoring the quantum entanglement between the different modes.  For example, the apparent source location for a particle emitted at late times will be roughly where the black hole was at the emission time, which depends on the history of the emission of momenta of the earlier particles that determines the black hole recoil that leads to the fluctuations in position.  When this quantum entanglement between modes is taken into account (which is absent in the semiclassical approximation that has the black hole remain at the fixed location of the center of mass of the total system, that of the original black hole before any radiation was emitted, and decay with a fixed $M(t)$ that just depends on the expectation value of the emission rate with no quantum fluctuations of the black hole in the semiclassical approximation), I would expect that the von Neumann entropy of the radiation is close to that of the semiclassical approximation up to the time $t_\ast$ at which the semiclassical entropy of the radiation reaches the Bekenstein-Hawking entropy of the black hole, which is presumably a good approximation for the maximum von Neumann entropy of the black hole that always stays equal to the von Neumann entropy of the radiation under my assumptions of a pure initial state, unitary evolution, and the possibility to divide the total system up into black hole and radiation subsystems.

Now the stochastic recoil of the black hole might be expected to increase its von Neumann entropy (and hence also that of the Hawking radiation), but this appears to be by an amount that is quite negligible compared with the Bekenstein-Hawking entropy $\sim M_0^2$.  In particular, since the black hole develops a position uncertainty $\sim M_0^2$ from its momentum uncertainty $\sim 1$ (all in Planck units) from the square root of $N \sim M_0^2$ particles emitted each with momentum $\sim 1/M_0$, the number of momentum states up to momentum $\sim 1$ in the region of volume $\sim M_0^6$ would be of the order of $\sim M_0^6$, which would lead to an entropy of the order of the logarithm of this number of states, or $\sim 6\ln{M_0} \sim 522 + 6\ln{(M_0/M_\odot)}$.  Although this is a large compared with unity, and would lead to a number of states far larger than a googol raised to the nine-fourths power, it is miniscule compared with the maximum von Neumann entropy of the radiation that is here estimated to be $S_\ast \approx 6.268\times 10^{76} (M_0/M_\odot)^2$, and also very much smaller than the uncertainty of this that I have estimated to be $\sim M_0 \sim 10^{38} (M_0/M_\odot)$.

Let us now consider the (rather hypothetical) possibility that the black hole forms with initial von Neumann entropy $S_0 = f\tilde{S}_\mathrm{BH}(0) = f(4\pi M_0^2)$ that is a significant fraction $f$ of the initial Bekenstein-Hawking thermodynamic entropy $\tilde{S}_\mathrm{BH}(0) = 4\pi M_0^2$. Let us suppose that the black hole system forms initially maximally entangled with a reference system $(X)$ \cite{Braunstein} that is thereafter assumed not to interact significantly with the rest of the universe.  For brevity, let the black hole system be $(Y)$ and the Hawking radiation system be $(Z)$, and assume that the entire $(XYZ)$ system is in a pure state, so that its von Neumann entropy is $S(XYZ) = 0$.  Let the effective Hilbert-space dimensions of these three systems be $X = \exp{(S_0)}$, $Y = \exp{[\tilde{S}_\mathrm{BH}(t)]}$, and $Z = \exp{[\tilde{S}_\mathrm{rad}(t)]}$, with the latter two changing with time.  We want the case in which the reference system $(X)$ is maximally entangled with the black hole plus Hawking radiation system $(YZ)$ in a total pure state, so that $S(YZ) = S(X) = \ln{X} = S_0 = f\tilde{S}_\mathrm{BH}(0) = f(4\pi M_0^2)$.

Since the reference system $(X)$ always has a Hilbert-space dimension not greater than that of the remaining $(YZ)$ system, $X \leq YZ$, my 1993 results \cite{Page:1993df,Page:1993wv} imply that for a Haar-measure random pure state of the entire system, one would have $S(X) = S(YZ) \approx \ln{X} = S_0$, which is what we want (nearly maximal entanglement between the reference system $(X)$ and the black hole plus Hawking radiation system $(YZ)$), so with negligible error we can assume that the quantum state of the entire system is a random pure state.  Then any subsystem with Hilbert-space dimension less than that of the remainder of the pure system would have nearly maximal von Neumann entropy, the logarithm of the dimension of that subsystem (for its given time-dependent effective Hilbert-space dimension).  These considerations lead to three stages in the evolution of the hole-radiation system $(YZ)$:

(1)  When $XZ \leq Y$ (which also implies $Z < XY$ for $X > 1$), one has that both $(XZ)$ and $(Z)$ have nearly maximum von Neumann entropy, so then the von Neumann entropy of the black hole is $S(Y) = S(XZ) \approx \ln{XZ} \approx S_0 + \tilde{S}_\mathrm{rad}(t)$ (not generically maximized at $\tilde{S}_\mathrm{BH}(t)$), and the von Neumann entropy of the Hawking radiation is $S_\mathrm{vN}(t) \equiv S(Z) = S(XY) \approx \ln{Z} = \tilde{S}_\mathrm{rad}(t)$ (very nearly maximized at the logarithm of the effective Hilbert-space dimension of the radiation at that time).

(2)  When $Y \leq XZ \leq X^2Y$, one has that both $(Y)$ and $(Z)$ have nearly maximum von Neumann entropies (for the effective Hilbert-space dimensions $Y$ and $Z$ at that time), so the von Neumann entropy of the black hole is $S(Y) = S(XZ) \approx \ln{Y} = \tilde{S}_\mathrm{BH}(t)$ (maximized), and the von Neumann entropy of the Hawking radiation is $S_\mathrm{vN}(t) \equiv S(Z) = S(XY) \approx \ln{Z} = \tilde{S}_\mathrm{rad}(t)$ (also maximized).

(3)  When $XY \leq Z$ (which also implies $Y < XZ$ for $X > 1$), one has that both $(XY)$ and $(Y)$ have nearly maximal von Neumann entropies, so then the von Neumann entropy of the black hole is $S(Y) = S(XZ) \approx \ln{Y} = \tilde{S}_\mathrm{BH}(t)$ (maximized), and the von Neumann entropy of the Hawking radiation is $S_\mathrm{vN}(t) \equiv S(Z) = S(XY) \approx \ln{X} + \ln{Y} = S_0 + \tilde{S}_\mathrm{BH}(t)$ (not generically maximized at $\tilde{S}_\mathrm{rad}(t)$).

It is useful to define a reparametrized retarded time coordinate $\tau$ that is the fraction of the Bekenstein-Hawking thermodynamic entropy $A/4$ that has been lost during the black hole evaporation,
\begin{equation}
\begin{split}
\tau &\equiv 1 - \tilde{S}_\mathrm{BH}(t)/\tilde{S}_\mathrm{BH}(0) \\
     &= 1 - A/A_0 = 1 - M^2/M_0^2 \\
     &= 1 - (1 - t/t_\mathrm{decay})^{2/3}.
\label{tau-time}
\end{split}
\end{equation}
Then the thermodynamic entropies have the form $\tilde{S}(X) = S(X) = \ln{X} = S_0 = f \tilde{S}_\mathrm{BH}(0)$, $\tilde{S}(Y) = \tilde{S}_\mathrm{BH}(t) = \ln{Y} = (1-\tau)\tilde{S}_\mathrm{BH}(0)$, and $\tilde{S}(Z) = S_\mathrm{vN}(t) = \ln{Z} = \beta\tau$.

Therefore, in Stage (1) of the evaporation, $XZ \leq Y$ implies $f + \beta\tau \leq 1-\tau$, which implies $\tau \leq \tau_{12} = (1-f)/(1+\beta)$.  During this stage, which is for
\begin{equation}
0 \leq \tau \leq \tau_{12} = \frac{1-f}{1+\beta},\ 0 \leq t \leq t_{12} 
  = \left[1-\left(\frac{\beta+f}{\beta+1}\right)^{3/2}\right]t_\mathrm{decay},
\label{tau1}
\end{equation}
the von Neumann entropy of the black hole is
\begin{equation}
S(Y) \approx \tilde{S}_\mathrm{BH}(0)(f+\beta\tau)
     = \tilde{S}_\mathrm{BH}(0)\left[f+\beta-\beta\left(1-\frac{t}{t_\mathrm{decay}}
           \right)^{2/3}\right],
\label{S1Y}
\end{equation}
the von Neumann entropy of the Hawking radiation is
\begin{equation}
S(Z) \approx \tilde{S}_\mathrm{rad}(t) = \tilde{S}_\mathrm{BH}(0)\beta\tau
     = \tilde{S}_\mathrm{BH}(0)\left[\beta-\beta\left(1-\frac{t}{t_\mathrm{decay}}
           \right)^{2/3}\right],
\label{S1Z}
\end{equation}
and the mutual information of the reference system $(X)$ and the Hawking radiation $(Z)$ is
\begin{equation}
I(X;Z) \equiv S(X)+S(Z)-S(XZ) \approx 0.
\label{mutual1}
\end{equation}

In Stage (2), $Y \leq XZ \leq X^2Y$ implies $1-\tau \leq f+\beta\tau$ and $\beta\tau \leq f+1-\tau$, which implies $\tau_{12} \leq \tau \leq \tau_{23} = (1+f)/(1+\beta)$.  During this stage, which is for
\begin{equation}
\tau_{12} \leq \tau \leq \tau_{23} = \frac{1+f}{1+\beta},\ t_{12} \leq t \leq t_{23}
  = \left[1-\left(\frac{\beta-f}{\beta+1}\right)^{3/2}\right]t_\mathrm{decay},
\label{tau2}
\end{equation}
the von Neumann entropy of the black hole is
\begin{equation}
S(Y) \approx \tilde{S}_\mathrm{BH}(t) = \tilde{S}_\mathrm{BH}(0)(1-\tau)
     = \tilde{S}_\mathrm{BH}(0)\left(1-\frac{t}{t_\mathrm{decay}}\right)^{2/3},
\label{S2Y}
\end{equation}
the von Neumann entropy of the Hawking radiation is
\begin{equation}
S(Z) \approx \tilde{S}_\mathrm{rad}(t) = \tilde{S}_\mathrm{BH}(0)\beta\tau
     = \tilde{S}_\mathrm{BH}(0)\left[\beta-\beta\left(1-\frac{t}{t_\mathrm{decay}}\right)^{2/3}\right],
\label{S2Z}
\end{equation}  
and the mutual information of the reference system $(X)$ and the Hawking radiation $(Z)$ is
\begin{equation}
I(X;Z) \approx 2S_0\frac{\tau-\tau_{12}}{\tau_{23}-\tau_{12}}
= 2S_0\frac{(t_\mathrm{decay}-t_{12})^{2/3}-(t_\mathrm{decay}-t)^{2/3}}
           {(t_\mathrm{decay}-t_{12})^{2/3}-(t_\mathrm{decay}-t_{23})^{2/3}}.
\label{mutual2}
\end{equation}

In Stage 3, $XY \leq Z$ implies $f+1-\tau \leq \beta\tau$, which implies $\tau_{23} \leq \tau \leq 1$.  During this stage, which is for
\begin{equation}
\tau_{23} \leq \tau \leq 1,\ t_{23} \leq t \leq t_\mathrm{decay} \approx 8895 M_0^3,
\label{tau3}
\end{equation}
the von Neumann entropy of the black hole is
\begin{equation}
S(Y) \approx \tilde{S}_\mathrm{BH}(t) = \tilde{S}_\mathrm{BH}(0)(1-\tau)
     = \tilde{S}_\mathrm{BH}(0)\left(1-\frac{t}{t_\mathrm{decay}}\right)^{2/3},
\label{S3Y}
\end{equation}
the von Neumann entropy of the Hawking radiation is
\begin{equation}
S(Z) \!\approx\! S_0 + \tilde{S}_\mathrm{BH}(t) \!=\! \tilde{S}_\mathrm{BH}(0)(f+1-\tau)
     \!=\! \tilde{S}_\mathrm{BH}(0)
     \left[f+\left(1-\frac{t}{t_\mathrm{decay}}\right)^{2/3}\right],
\label{S3Z}
\end{equation}  
and the mutual information of the reference system $(X)$ and the Hawking radiation $(Z)$ is
\begin{equation}
I(X;Z) \equiv S(X)+S(Z)-S(XZ) \approx 2S_0,
\label{mutual3}
\end{equation}
which is its maximal possible value.

\newpage

We can now use the Heaviside step function $\theta(x)$ to give a single formula, using Planck units, for the full time dependence (from the initial black hole formation at $\tau = 0$ or $t=0$ to its complete evaporation at $\tau = 1$ or $t = t_\mathrm{decay} \approx 8895 M_0^3$) of the von Neumann entropy of the black hole whose initial von Neumann entropy $S_\mathrm{BH}(0)$ is the fraction $f$ of the initial Bekenstein-Hawking entropy $\tilde{S}_\mathrm{BH}(0) = 4\pi M_0^2$:
\begin{equation}
\begin{split}
S_\mathrm{BH} \approx {} &
\tilde{S}_\mathrm{BH}(0)
\left[\theta(\tau_{12}-\tau)(f+\beta\tau) + \theta(\tau-\tau_{12})(1-\tau)\right]
\\
= {} &
4\pi M_0^2
\Biggl\{\theta\left(t_{12}-t\right)
\left[f+\beta-\beta\left(1-\frac{t}{t_\mathrm{decay}}\right)^{2/3}\right]
\\
& \hspace{15mm} +\theta\left(t-t_{12}\right)
\left(1-\frac{t}{t_\mathrm{decay}}\right)^{2/3}\Biggr\}
\\
\approx {} &
4\pi M_0^2 \Biggl\{\theta
\left[
8895 M_0^3\left(1-\left\{\frac{1.4847+f}{2.4847}\right\}^{3/2}\right)-t\right]
\times
\\
& \hspace{19mm}
\left[f+1.4847\left(1-\left\{1-\frac{t}{8895 M_0^3}\right\}^{2/3}\right)\right]
\\
& \hspace{10mm}
+\theta\left[t-8895 M_0^3
\left(1-\left\{\frac{1.4847+f}{2.4847}\right\}^{3/2}\right)\right] \times
\\
& \hspace{48mm}
\left(1-\frac{t}{8895 M_0^3}\right)^{2/3}\Biggr\}.
\label{impure-BH-von-Neumann-entropy-evolution}
\end{split}
\end{equation}
At $\tau=\tau_{12}$ or $t_{12} = t_\mathrm{decay}\{1 - [(\beta+f)/(1+\beta)]^{3/2}\}$, the von Neumann entropy of the black hole reaches its peak of $S_\mathrm{BH}(t_{12}) = [(f+\beta)/(1+\beta)]\tilde{S}_\mathrm{BH}(0)$, which is less than the initial Bekenstein-Hawking entropy $\tilde{S}_\mathrm{BH}(0) = 4\pi M_0^2$ unless the black hole starts maximally mixed, $S_\mathrm{BH}(0) = \tilde{S}_\mathrm{BH}(0)$, or $f=1$.

\newpage

Similarly, one can write a single formula for the full time dependence of the von Neumann entropy of the Hawking radiation as
\begin{equation}
\begin{split}
S_\mathrm{vN} \approx {} &
\tilde{S}_\mathrm{BH}(0)
[\theta(\tau_{23}-\tau)(\beta\tau) + \theta(\tau-\tau_{23})(1+f-\tau)]
\\
= {} &
4\pi M_0^2
\Biggl\{\theta(t_{23}-t)\left[\beta
-\beta\left(1-\frac{t}{t_\mathrm{decay}}\right)^{2/3}\right]
\\
& \hspace{15mm} +\theta(t-t_{23})\left[f+\left(1-\frac{t}{t_\mathrm{decay}}\right)^{2/3}
\right]\Biggr\}
\\
\approx {} &
4\pi M_0^2 \Biggl\{\theta
\left[
8895 M_0^3\left(1-\left\{\frac{1.4847-f}{2.4847}\right\}^{3/2}\right)-t\right]
\times
\\
& \hspace{22mm}
1.4847\left[1-\left(1-\frac{t}{8895 M_0^3}\right)^{2/3}\right]
\\
& \hspace{15mm}
+\theta\left[t-8895 M_0^3
\left(1-\left\{\frac{1.4847-f}{2.4847}\right\}^{3/2}\right)\right] \times
\\
& \hspace{35mm}
\left[f+\left(1-\frac{t}{8895 M_0^3}\right)^{2/3}\right]\Biggr\}.
\label{impure-rad-von-Neumann-entropy-evolution}
\end{split}
\end{equation}
At $\tau=\tau_{23}$ or $t_{23} = t_\mathrm{decay}\{1 - [(\beta-f)/(1+\beta)]^{3/2}\}$, the von Neumann entropy of the Hawking radiation reaches its peak of $S_\mathrm{BH}(t_{23}) = \beta[(1+f)/(1+\beta)]\tilde{S}_\mathrm{BH}(0)$,
which for $f = 1$ is $[2\beta/(1+\beta)]\tilde{S}_\mathrm{BH}(0) \approx 1.1951\,\tilde{S}_\mathrm{BH}(0)$ at $t_{23} \approx 0.9138\,t_\mathrm{decay}$.

One can also write a single formula for the mutual information of the reference system $(X)$ and the Hawking radiation $(Z)$ as

\begin{equation}
I(X;Z) \approx 2S_0\theta\left(\tau-\tau_{12}\right)
\left[\theta\left(\tau_{23}-\tau\right)
\frac{\tau-\tau_{12}}{\tau_{23}-\tau_{12}}
+\theta\left(\tau-\tau_{23}\right)\right].
\label{mutual-information-evolution}
\end{equation}

Plots of the von Neumann entropy of the black hole and of the Hawking radiation, as well as the mutual information of the reference system and the Hawking radiation, are given in the next ten pages.

\newpage

\begin{figure}[H]
\centering
\includegraphics[width=1\textwidth]{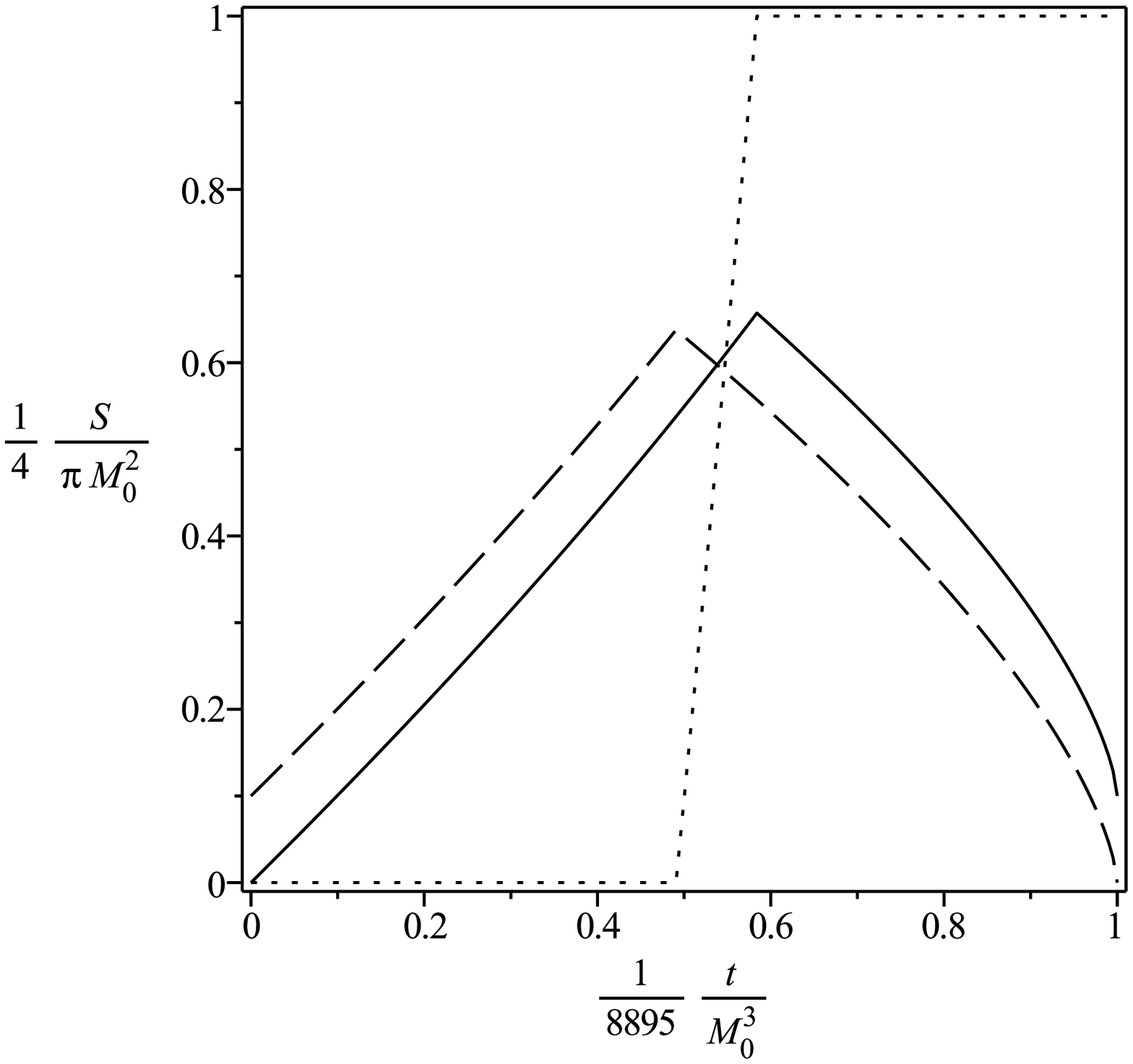}
\caption{Plot of Hole and Radiation Entropy vs.\ Time for $f=0.1$.
Solid line is the von Neumann entropy of the Hawking radiation.
Dashed line is the von Neumann entropy of the black hole.
Dotted line is the normalized mutual information of the reference system $(X)$ and the Hawking radiation $(Z)$, $[S(X)+S(Z)-S(XZ)]/[2S(X)]$.}
\end{figure}

\begin{figure}[H]
\centering
\includegraphics[width=1\textwidth]{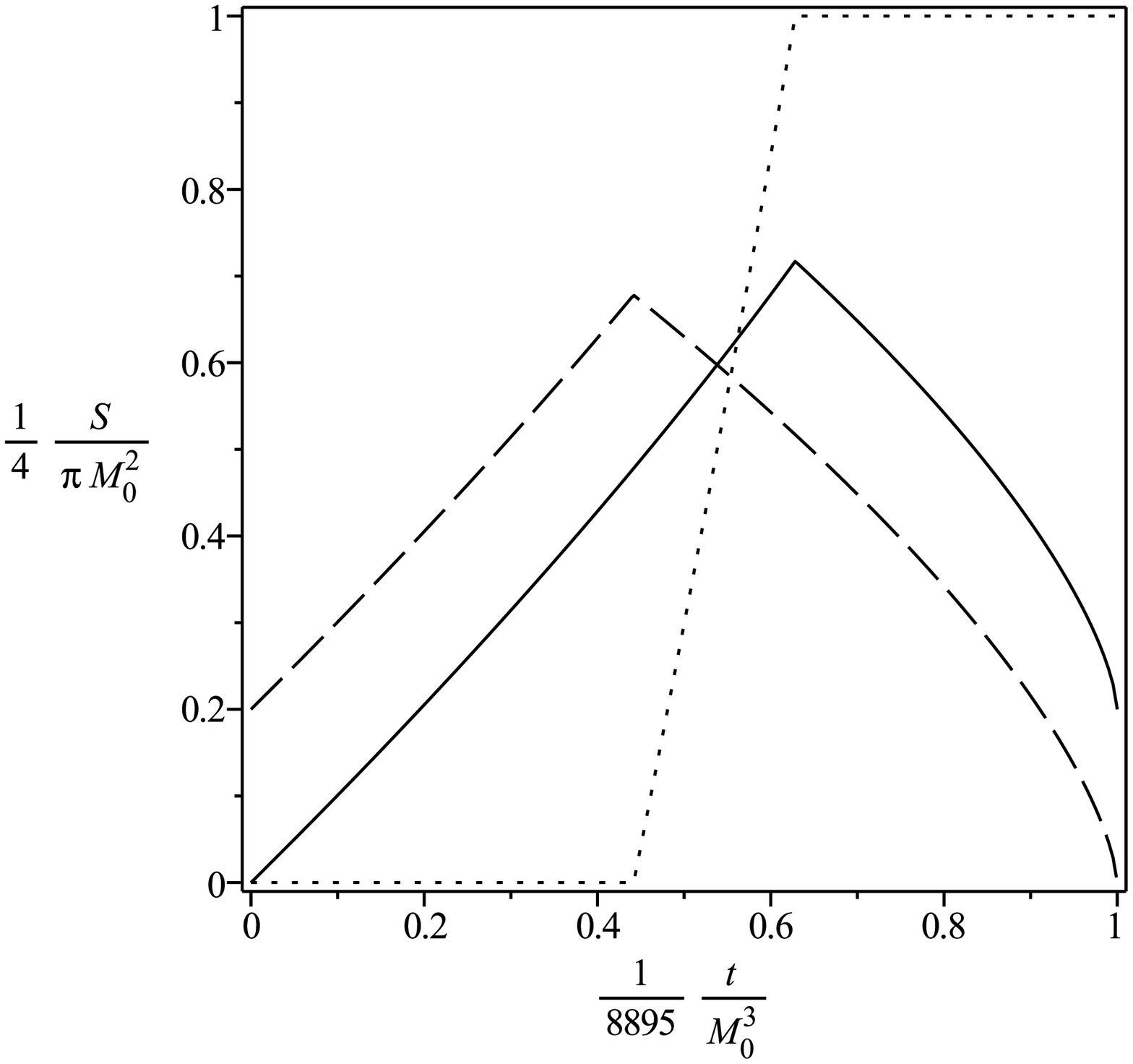}
\caption{Plot of Hole and Radiation Entropy vs.\ Time for $f=0.2$.
Solid line is the von Neumann entropy of the Hawking radiation.
Dashed line is the von Neumann entropy of the black hole.
Dotted line is the normalized mutual information of the reference system $(X)$ and the Hawking radiation $(Z)$, $[S(X)+S(Z)-S(XZ)]/[2S(X)]$.}
\end{figure}

\begin{figure}[H]
\centering
\includegraphics[width=1\textwidth]{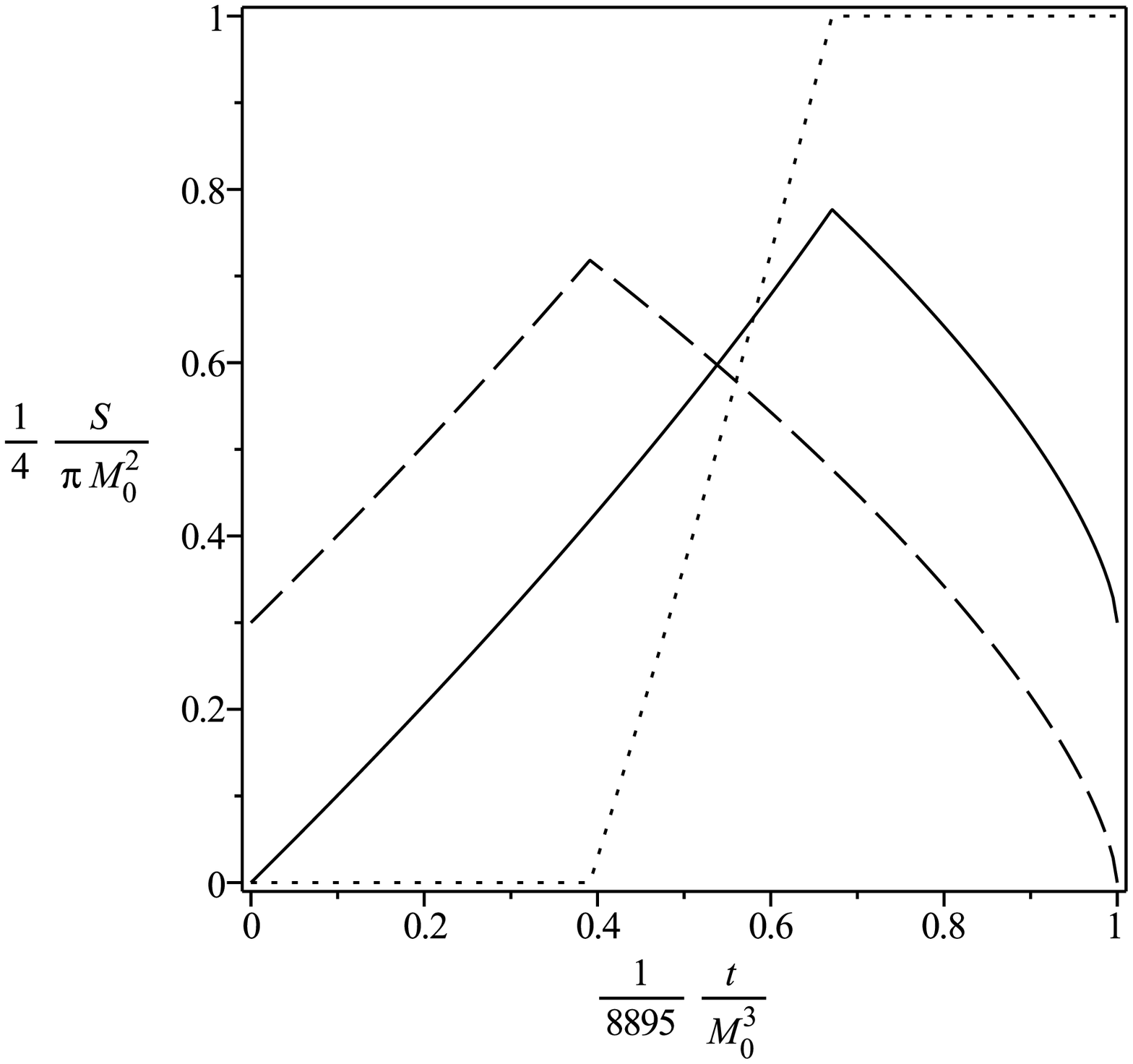}
\caption{Plot of Hole and Radiation Entropy vs.\ Time for $f=0.3$.
Solid line is the von Neumann entropy of the Hawking radiation.
Dashed line is the von Neumann entropy of the black hole.
Dotted line is the normalized mutual information of the reference system $(X)$ and the Hawking radiation $(Z)$, $[S(X)+S(Z)-S(XZ)]/[2S(X)]$.}
\end{figure}

\begin{figure}[H]
\centering
\includegraphics[width=1\textwidth]{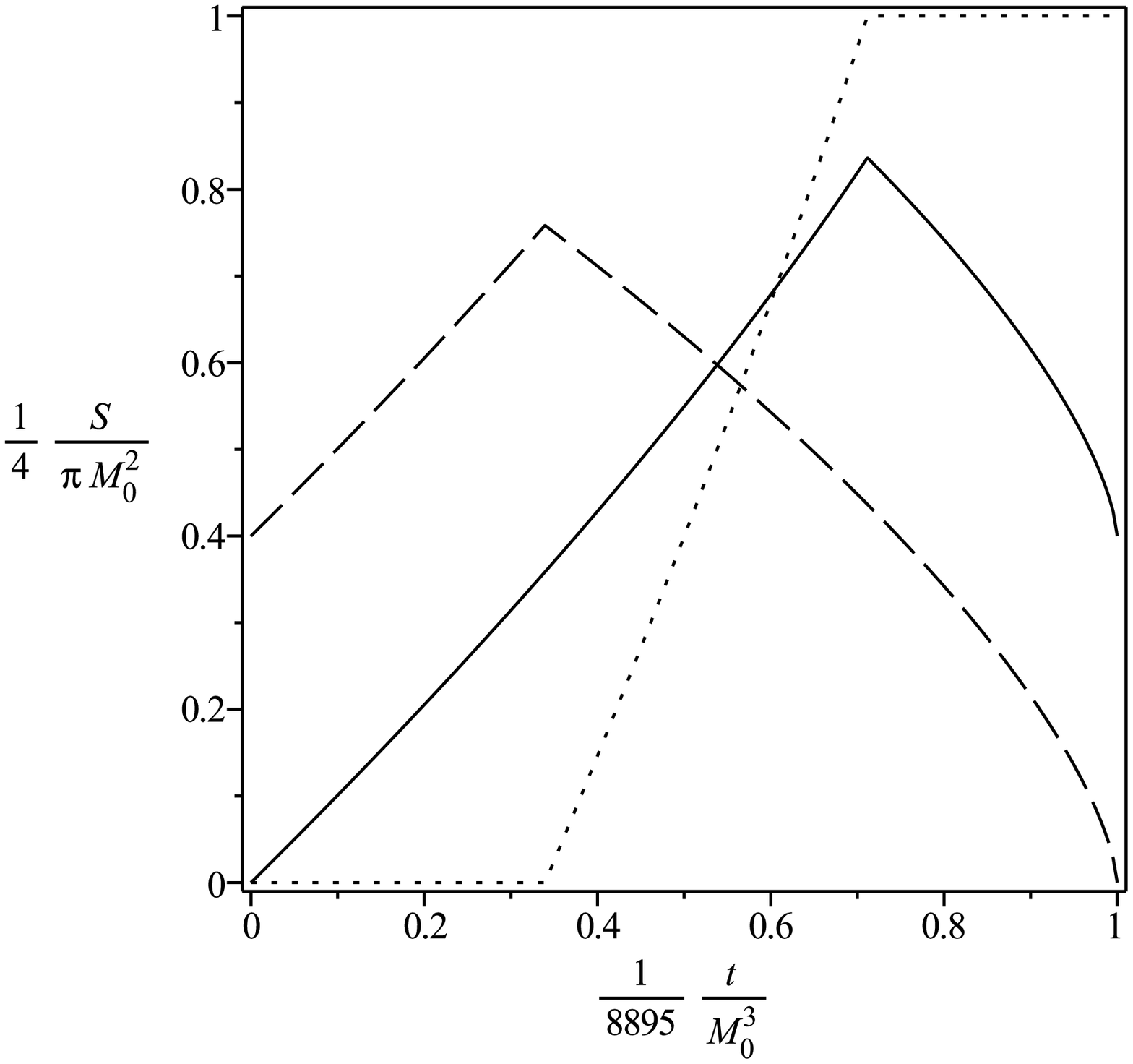}
\caption{Plot of Hole and Radiation Entropy vs.\ Time for $f=0.4$.
Solid line is the von Neumann entropy of the Hawking radiation.
Dashed line is the von Neumann entropy of the black hole.
Dotted line is the normalized mutual information of the reference system $(X)$ and the Hawking radiation $(Z)$, $[S(X)+S(Z)-S(XZ)]/[2S(X)]$.}
\end{figure}

\begin{figure}[H]
\centering
\includegraphics[width=1\textwidth]{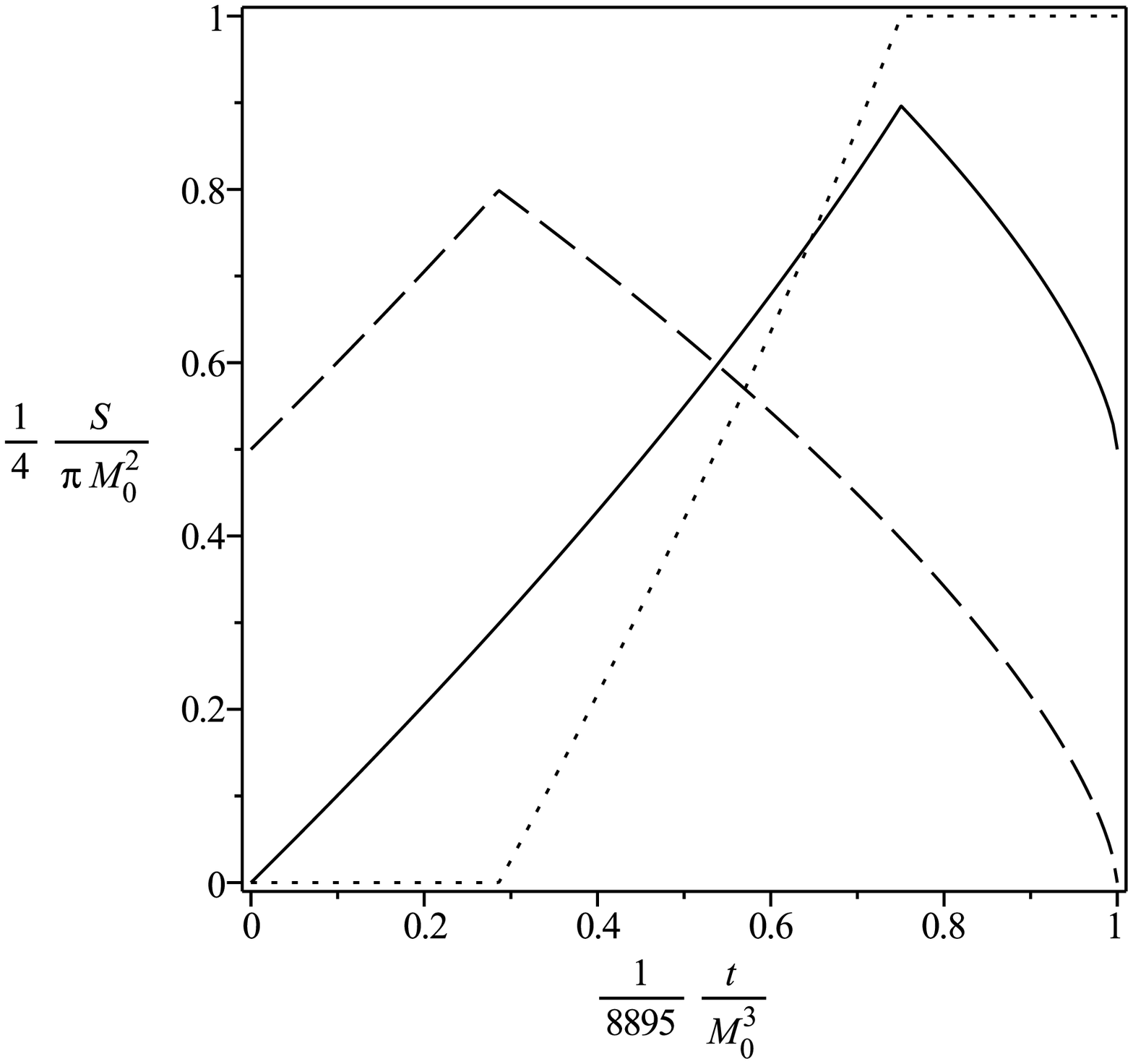}
\caption{Plot of Hole and Radiation Entropy vs.\ Time for $f=0.5$.
Solid line is the von Neumann entropy of the Hawking radiation.
Dashed line is the von Neumann entropy of the black hole.
Dotted line is the normalized mutual information of the reference system $(X)$ and the Hawking radiation $(Z)$, $[S(X)+S(Z)-S(XZ)]/[2S(X)]$.}
\end{figure}

\begin{figure}[H]
\centering
\includegraphics[width=1\textwidth]{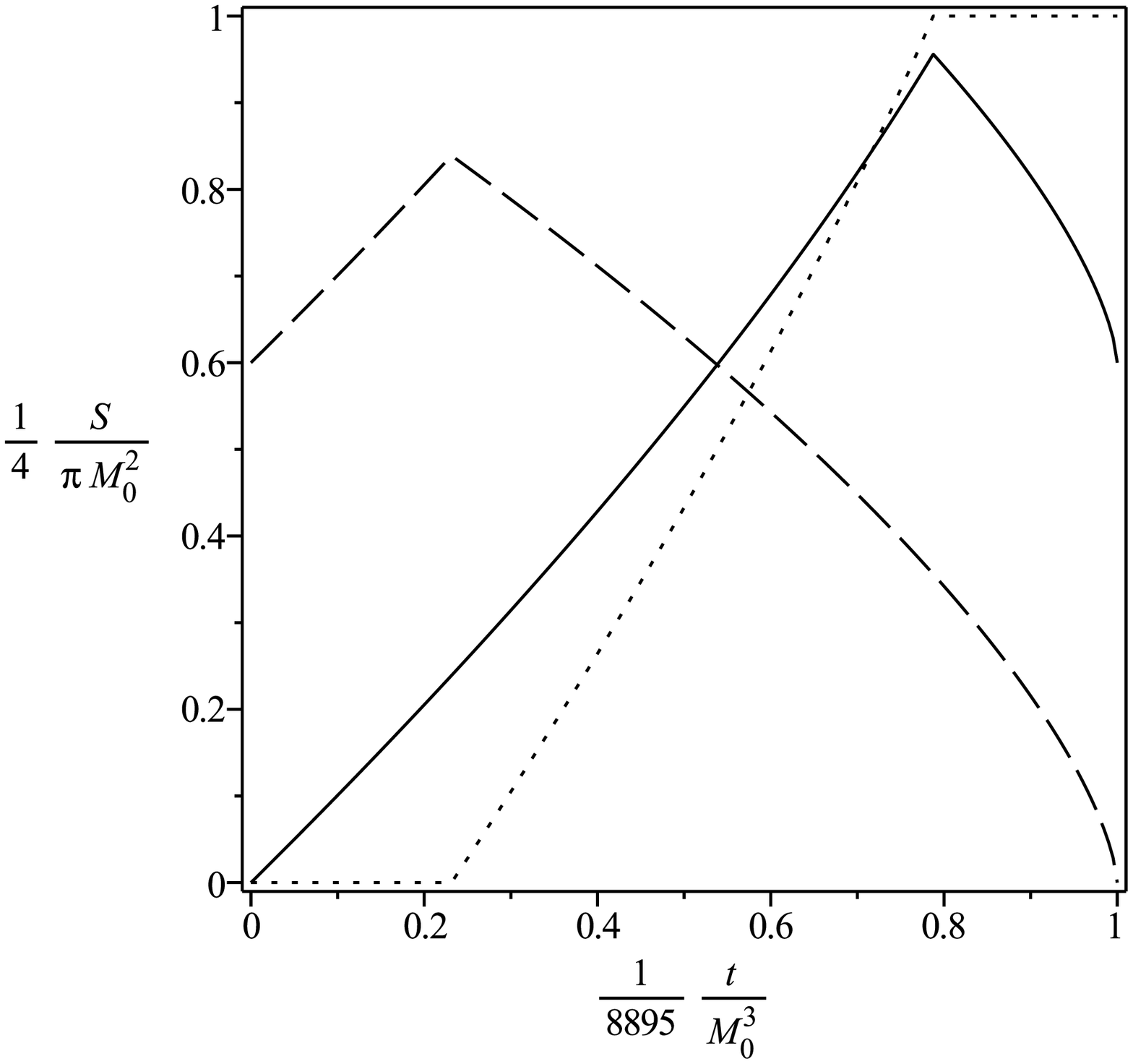}
\caption{Plot of Hole and Radiation Entropy vs.\ Time for $f=0.6$.
Solid line is the von Neumann entropy of the Hawking radiation.
Dashed line is the von Neumann entropy of the black hole.
Dotted line is the normalized mutual information of the reference system $(X)$ and the Hawking radiation $(Z)$, $[S(X)+S(Z)-S(XZ)]/[2S(X)]$.}
\end{figure}

\begin{figure}[H]
\centering
\includegraphics[width=1\textwidth]{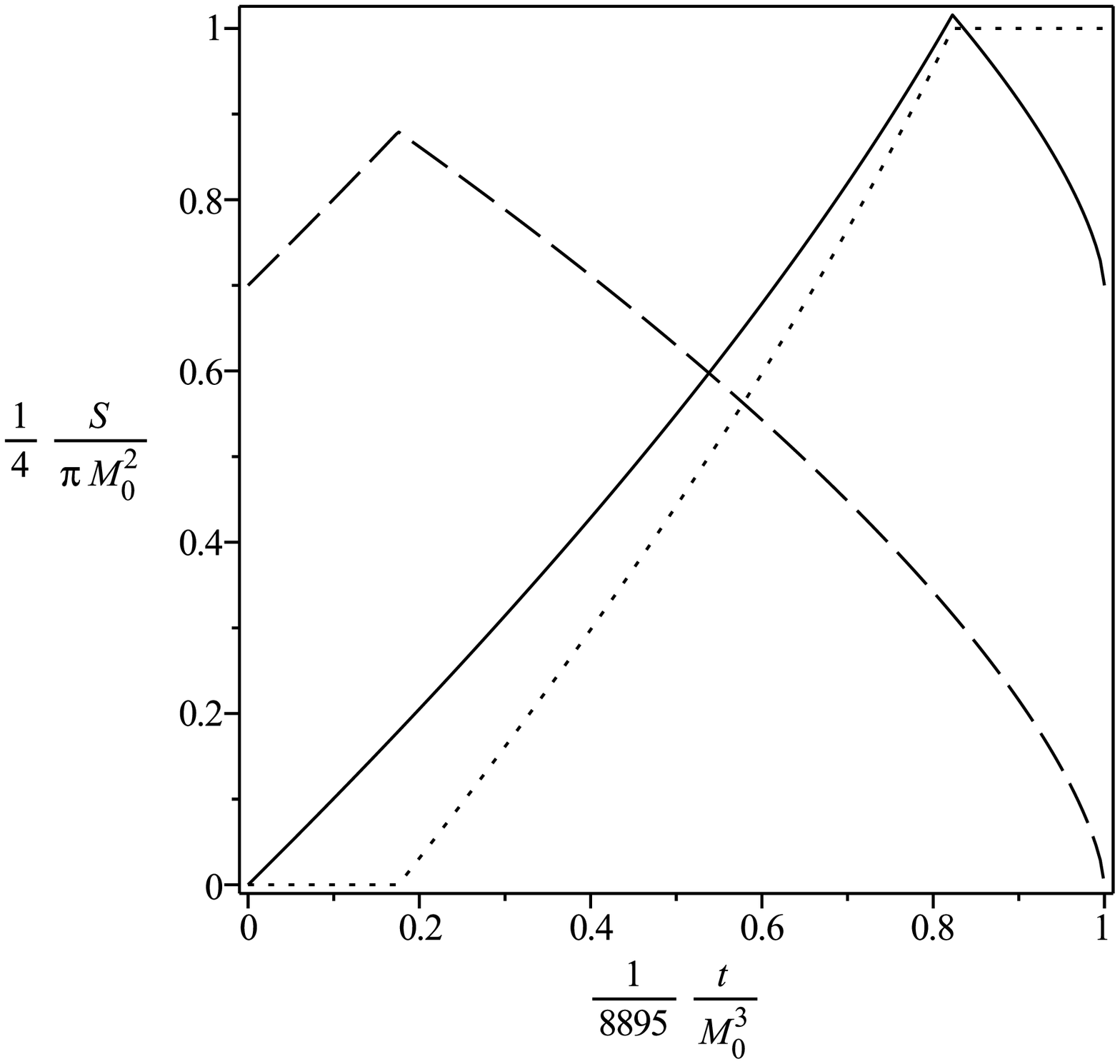}
\caption{Plot of Hole and Radiation Entropy vs.\ Time for $f=0.7$.
Solid line is the von Neumann entropy of the Hawking radiation.
Dashed line is the von Neumann entropy of the black hole.
Dotted line is the normalized mutual information of the reference system $(X)$ and the Hawking radiation $(Z)$, $[S(X)+S(Z)-S(XZ)]/[2S(X)]$.}
\end{figure}

\begin{figure}[H]
\centering
\includegraphics[width=1\textwidth]{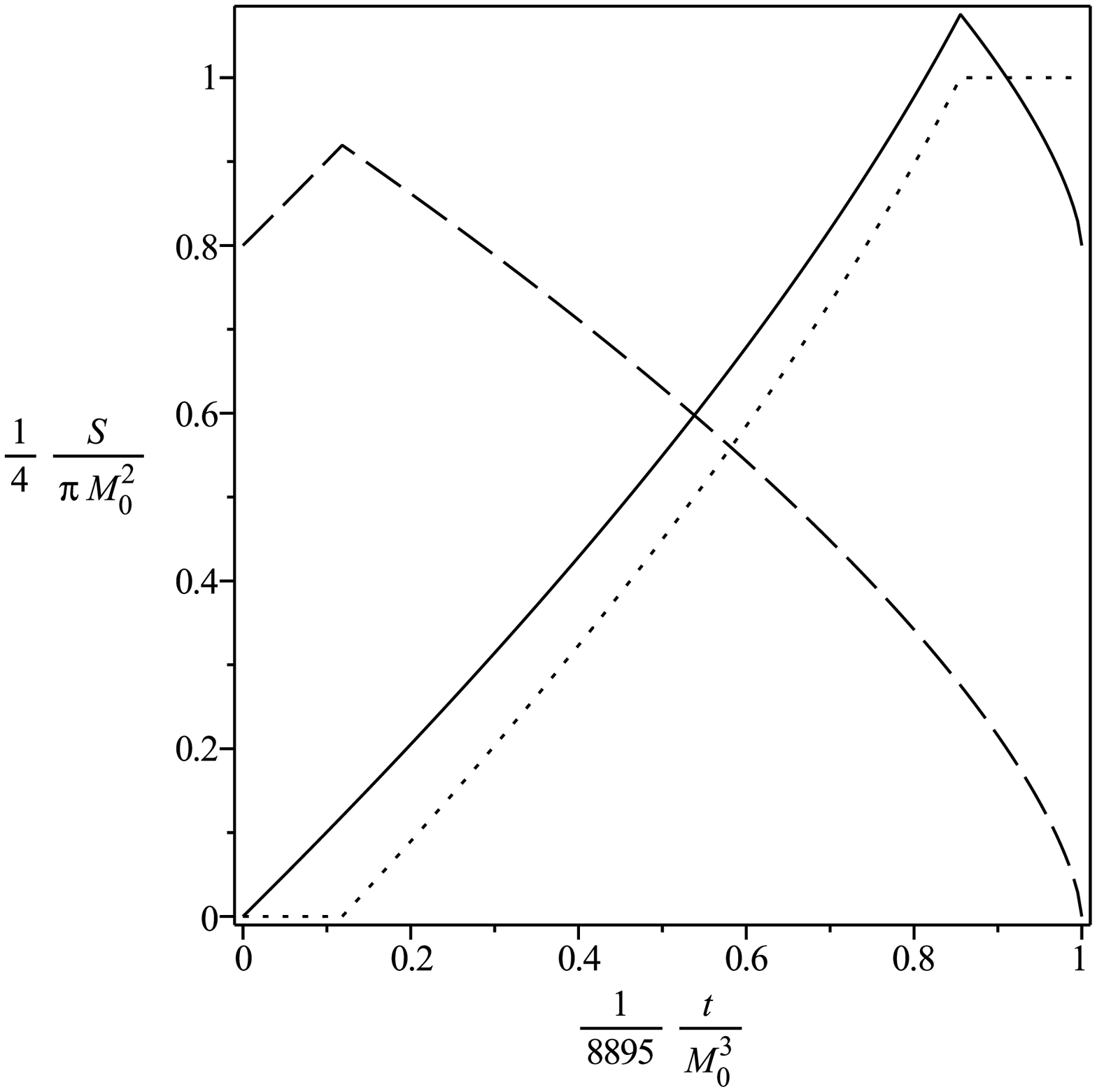}
\caption{Plot of Hole and Radiation Entropy vs.\ Time for $f=0.8$.
Solid line is the von Neumann entropy of the Hawking radiation.
Dashed line is the von Neumann entropy of the black hole.
Dotted line is the normalized mutual information of the reference system $(X)$ and the Hawking radiation $(Z)$, $[S(X)+S(Z)-S(XZ)]/[2S(X)]$.}
\end{figure}

\begin{figure}[H]
\centering
\includegraphics[width=1\textwidth]{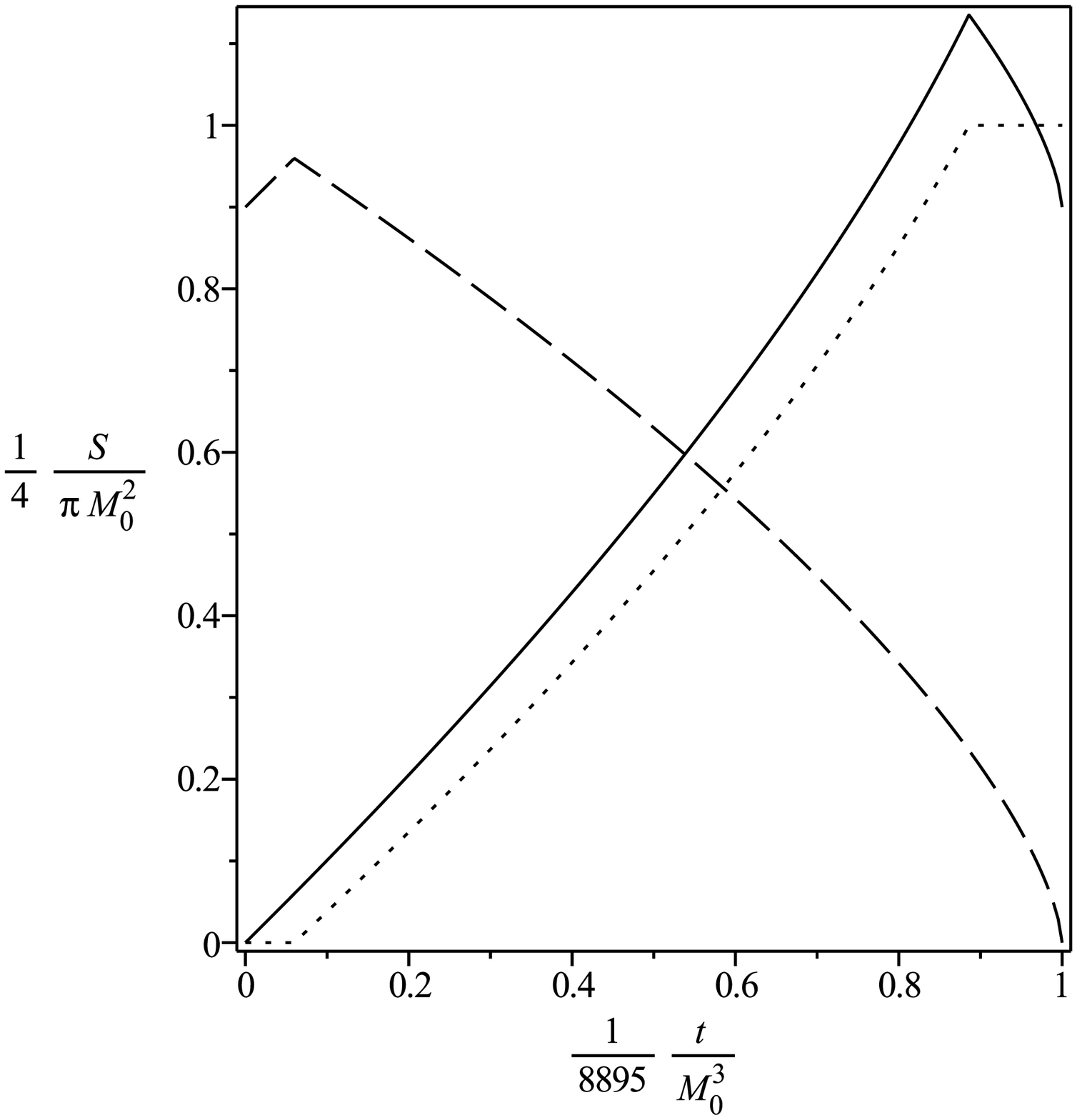}
\caption{Plot of Hole and Radiation Entropy vs.\ Time for $f=0.9$.
Solid line is the von Neumann entropy of the Hawking radiation.
Dashed line is the von Neumann entropy of the black hole.
Dotted line is the normalized mutual information of the reference system $(X)$ and the Hawking radiation $(Z)$, $[S(X)+S(Z)-S(XZ)]/[2S(X)]$.}
\end{figure}

\begin{figure}[H]
\centering
\includegraphics[width=1\textwidth]{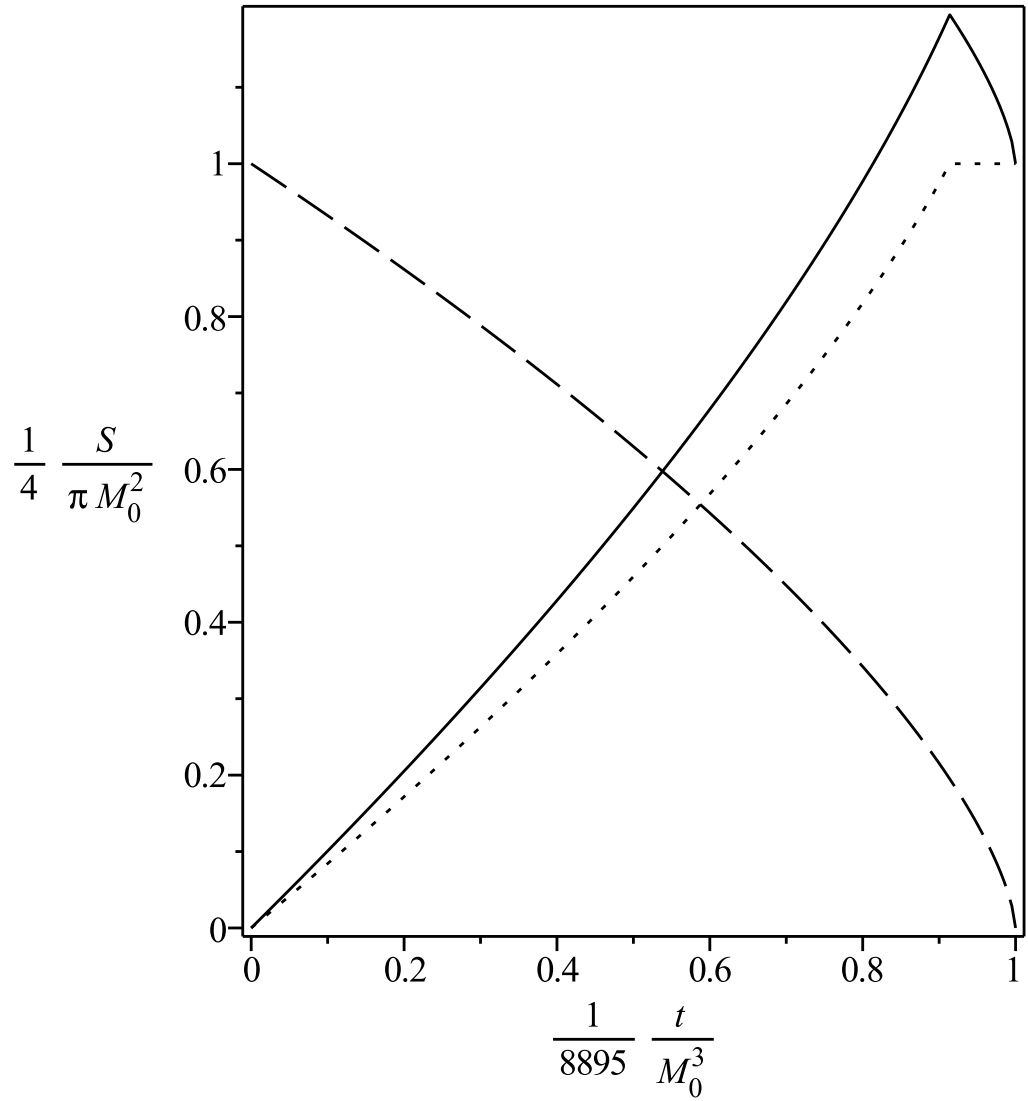}
\caption{Plot of Hole and Radiation Entropy vs.\ Time for $f=1.0$.
Solid line is the von Neumann entropy of the Hawking radiation.
Dashed line is the von Neumann entropy of the black hole.
Dotted line is the normalized mutual information of the reference system $(X)$ and the Hawking radiation $(Z)$, $[S(X)+S(Z)-S(XZ)]/[2S(X)]$.}
\end{figure}

\newpage

\baselineskip 19pt

In conclusion, under the assumptions that a large (say solar mass or greater) nonrotating uncharged (Schwarzschild) black hole starts in nearly a pure quantum state and decays away completely by a unitary process while being nearly maximally scrambled at all times, the von Neumann entropy of the Hawking radiation (almost entirely photons and gravitons for such a large black hole, assuming no other particles of rest mass $m \stackrel{<}{\sim} 10^{-10}$ eV) increases up to a maximum of $S_\ast \approx 0.59754(4\pi M_0^2) \approx 7.5089 M_0^2 \approx 6.268\times 10^{76} (M_0/M_\odot)^2$ at time $t_\ast \approx 0.53810\, t_\mathrm{decay} \approx 4786 M_0^3 \approx 6.236\times 10^{66}(M_0/M_\odot)^3\mathrm{yr}$ and then decreases back down to near zero (the von Neumann entropy of the original black hole, assumed to be small).  The numerically estimated time dependence of the von Neumann entropy of the Hawking radiation is given by Eq.\ (\ref{rad-von-Neumann-entropy-evolution}) and is plotted in Figure 1.

On the other hand, if the black hole starts in a mixed state with von Neumann entropy $f$ times its Bekenstein-Hawking thermodynamic entropy $4\pi M_0^2$, the time dependence of the von Neumann entropy of the black hole is given by Eq.\ (\ref{impure-BH-von-Neumann-entropy-evolution}), and that of the Hawking radiation is given by Eq.\ (\ref{impure-rad-von-Neumann-entropy-evolution}).  These quantities are graphed in Figures 2-11.

Appreciation is expressed to Leonard Susskind for inviting me to the 2012 November 30 - December 1 firewall conference at Stanford University, and to the participants there (particular Andrew Strominger, who asked explicitly about the time dependence of the radiated entropy), who motivated me to dig out my old numerical calculations to give the estimate of the time dependence of the von Neumannn entropy of the Hawking radiation.  This work was supported in part by the Natural Sciences and Engineering Research Council of Canada.

Revisions of this paper were made at the Cook's Branch Nature Conservancy, where I have greatly appreciated the hospitality of the Mitchell family and of the George P. and Cynthia W. Mitchell Institute for Fundamental Physics and Astronomy of Texas A \& M University.  This work was supported in part by the Natural Sciences and Engineering Council of Canada.


\begin{thebibliography}{99}
\bibitem{Cathy}
  I thank my wife Cathy Page for asking me whether a skinny triangle nearly saturating the triangular inequalities could be called anorexic, enabling me to name the CATH assumption in her honour.
\end{thebibliography}
\end{document}